\documentclass[aps,showpacs, nofootinbib,floatfix]{revtex4}
\usepackage{graphicx, graphics, color}
\usepackage{subfig} 

\input amssym.tex
\input amssym.def

\newcommand{\be}{\begin{equation}}
\newcommand{\ee}{\end{equation}}
\newcommand{\ben}{\begin{eqnarray}}
\newcommand{\een}{\end{eqnarray}}

\newcommand{\la}{{\lambda}}

\newcommand{\cA}{{\cal A}}
\newcommand{\cB}{{\cal B}}
\newcommand{\cC}{{\cal C}}

\newcommand{\cO}{{\cal O}}

\newcommand{\cP}{{\cal P}}
\newcommand{\cQ}{{\cal Q}}
\newcommand{\cR}{{\cal R}}

\newcommand{\p}{\partial}
\newcommand{\na}{\nabla}

\newcommand{\tr}{{\tilde \rho}}

\begin{document}

\title{Analytic investigation of holographic phase transitions influenced by dark matter sector}

\author{{\L}ukasz Nakonieczny}
\email{Lukasz.Nakonieczny@fuw.edu.pl}
\affiliation{Institute of Theoretical Physics, Faculty of Physics, University of Warsaw \protect \\
ul.~Pasteura 5,~02-093 Warszawa, Poland }

\author{Marek Rogatko} 
\email{rogat@kft.umcs.lublin.pl,
marek.rogatko@poczta.umcs.lublin.pl }
\author{Karol I. Wysoki\'nski}
\email{karol.wysokinski@umcs.pl}
\affiliation{Institute of Physics \protect \\
Maria Curie-Sklodowska University \protect \\
20-031 Lublin, pl.~Marii Curie-Sklodowskiej 1, Poland}

\date{\today}

\begin{abstract}
We analytically study the phase transitions between s-wave holographic
insulator/superconductor and metal/superconductor. The  problem is solved 
by the variational method for the Sturm-Liouville eigenvalue problem
in the theory with dark matter sector of U(1)-gauge field coupled to the Maxwell field.
Additionally in the probe limit we  investigate  the marginally stable modes of scalar perturbations
in the AdS solitonic background, connected with magnetic field in the dark matter sector.
We have found that even with dark matter sector the superconducting  transition temperature $T_c$ is proportional to  
charge density $\rho$ in power 1/3. This value seem to be strong coupling modification of the
exponent 2/3 known from the Bose - Einstein condensation of charged local pair bosons in narrow band superconductors. 
The holographic droplet solution is affected by the coupling to the dark matter. Interestingly in the probe
limit the critical chemical potential increases with the decreasing  coupling to dark matter
making the condensation transition harder to appear.
\end{abstract}

\maketitle

\section{Introduction}
The gauge/gravity duality in the form of AdS/CFT correspondence \cite{mal,wit} 
provides an interesting framework to study strong coupling  effects in quantum many body $d$-dimensional 
systems \cite{sachdev2012} by means of the $d+1$ dimensional spacetime with the negative cosmological
constant.  In particular, this technique has been widely used to describe phase  transitions from 
the normal or insulating to superconducting state. In the original work on 'building the holographic 
superconductor' \cite{har08,har08b}, the description of the single band s-wave superconductor was proposed.
The scalar complex field with the appropriate potential has been incorporated into
the theory of gravity and the condensation of its dual operator at finite temperature $T$, lower than the
critical one $T_c$  was observed. The condensed operator has been identified with the superconducting order parameter. 
In the bulk, the temperature was introduced as the Hawking's black hole one. 

The  aforementioned approach has been extended in 
many directions, taking into account the relevant aspects of the existing superconducting materials.
For instance, the models of $d$-wave \cite{d-wave} superconductivity have  been considered, to shed some light
on the strong coupling behaviour of the well known high temperature superconductors \cite{hts}, which 
feature this symmetry of the order parameter. The spin triplet superconducting  states of simple $p$-wave \cite{p-wave} 
as well as the chiral   $p_x \pm ip_y$ symmetry have also been  elaborated  in considerable details \cite{chiral-p}.  
In Refs.\cite{2bandsc} the  multi-band superconducting systems have  been studied in view of  many materials 
in which the coexistence of different orbitals plays a crucial role, e.g., in MgB$_2$ \cite{liu2001}, 
Sr$_2$RuO$_4$ \cite{mackenzie2003} or in heavy fermion superconductors  \cite{mathur1998}.  The  description of these  
 materials requires at least two hybridized orbitals which in the gravity approach translate into two scalar fields.

\par
On the other hand, apart from the aforementioned studies of conductor/superconductor phase transitions,
the holographic insulator/superconductor transitions attracted a great attention. Modification of 
the bulk gravity theory by considering the five-dimensional AdS soliton line element \cite{hor98}
coupled to Maxwell gauge field and scalar one, allows for building a model of holographic insulator/superconductor
phase transition at zero temperature \cite{nis10}. In gauge/gravity duality description,
the AdS soliton is dual to a confined field theory with a mass gap, mimicking  an insulator phase \cite{wit98a}.
It was revealed  that in the presence of a chemical potential in the solitonic background,
the insulator/metal transition is of the second order. Namely, for the chemical potential greater
than some critical value, the considered background turned out to be unstable and non-trivial
hair emerged. This fact is interpreted as insulator/superconductor phase transitions. 

In Refs.\cite{hor10,bri11} it was shown that the strength of various kinds of matter backreactions could 
generate new types of phase transitions. Marginally stable modes of scalar/vector perturbations in the AdS solitonic spacetime
were studied in \cite{cai11a,cai11b} to reveal the onset of the phase transition as well as to find the magnetic field
effects on them. Among others, it was claimed that magnetic field made  the phase transition
harder to occur. The compatibility with the earlier investigations \cite{gub08} were also announced.

Recently, the influence of nonlinear electrodynamics on the holographic insulator/superconductor phase transitions
was taken into account \cite{zha13,jin12}, while the problem of $p$-wave symmetry of the transitions in question was 
treated in Ref.\cite{cai11c,akh11}. On the other hand, the analytical investigations tackling the phase transitions
of this type, in Gauss-Bonnet gravity were discussed in \cite{pan11}.

Moreover, superconducting solutions in which the condensate was confined to a finite region and decayed
rapidly outside during conductor/ superconductor phase transitions, were examined. Both the vortex and the droplet
models were constructed for s-wave type of superconductors \cite{alb09}.
A non-Abelian droplet solutions emerging during insulator/superconductor phase transition in $p$-wave
and $p+ip$-wave symmetry were studied in Ref.\cite{roy13}. It has been shown that in the case of Gauss-Bonnet background
the coupling constant of the theory (like in \cite{pan11}) affects the transition in question. 

Furthermore, the question of the possible matter configurations naturally appears in AdS spacetime.
The problem of the strictly stationary Einstein-Maxwell spacetime with negative cosmological constant was
treated in \cite{shi12}, while the simply connected Einstein-Maxwell-axion-dilaton spacetime with negative cosmological
constant and arbitrary number of $U(1)$-gauge fields was examined in Ref.\cite{bak13}. 
It was revealed that the considered spacetimes could not allow for the existence of nontrivial
configurations of complex scalar fields or form fields.

Motivated by the above problems  as well as to provide continuity with our previous studies \cite{nak14,nak15},
we address here the problem of phase transitions among insulator/superconductor and
metal/superconductor for s-wave holographic superconductors in the theory
in which {\it dark matter} sector is coupled to the Standard Model. We shall look for
the imprints of {\it dark matter} sector in possible holographic experiments.

The importance of examinations of such kind of models goes back to the need of explanation of
$511$ keV gamma rays astrophysical observations made by Integral/SPI \cite{integral} as well as
the experiments showing the electron positron excess in galaxy, revealed by ATIC/PAMELA \cite{atic, pamela}. 
 Their energies vary from a few GeV to a few TeV depending on the  experiments. 
On the other hand, the new physics can explain the $3.6\sigma$ discrepancy between measured value 
of the muon anomalous magnetic moment and its prediction in the Standard Model \cite{muon}.
The other facet concerns the fact that dark matter model 
is subject to the key ingredient in early Universe, 
where the topological phase transition, giving rise to various topological defects, might have happened.

Our analysis will be addressed to the theory in which, apart from the gravitational action given by
\be
S_{g} = \int \sqrt{-g}~ d^5 x~  \frac{1}{2 \kappa^2}\bigg( R - 2\Lambda\bigg), 
\ee
where $\kappa^2 = 8 \pi G_{5}$ is five-dimensional gravitational constant, $\Lambda = - 6/ L^2$ stands for the cosmological
constant, while $L$ is the radius of the AdS spacetime, we shall examine the Abelian-Higgs sector coupled to the 
second $U(1)$-gauge field
\be
\label{s_matter}
S_{m} = \int \sqrt{-g}~ d^5x  \bigg( 
- \frac{1}{4}F_{\mu \nu} F^{\mu \nu} - \left [ \nabla_{\mu} \psi - 
i q A_{\mu} \psi \right ]^{\dagger} \left [ \nabla^{\mu} \psi - i q A^{\mu} \psi  \right ]
- V(\psi) - \frac{1}{4} B_{\mu \nu} B^{\mu \nu} - \frac{\alpha}{4} F_{\mu \nu} B^{\mu \nu}
\bigg), 
\ee  
where the scalar field potential satisfies $V(\psi) = m^2 |\psi|^2 + \frac{\lambda_{\psi}}{4} |\psi|^4$.
$F_{\mu \nu} = 2 \nabla_{[ \mu} A_{\nu ]}$ stands for the ordinary Maxwell field strength tensor, while
the second $U(1)$-gauge field $B_{\mu \nu}$ is given by $B_{\mu \nu} = 2 \nabla_{[ \mu} B_{\nu ]}$. 
Moreover, $m,~ q$ represent  a mass and a charge related to the scalar 
field $\psi$. Here $\alpha$ is a coupling constant between $U(1)$ fields. The compatibility
with the current observations establishes its order to $10^{-3}$.

Within the above model the backreaction problems of the {\it dark matter} sector on s-wave holographic
superconductor was analyzed in Ref. \cite{nak14}. It was revealed that the {\it dark matter} coupling constant is bigger
the smaller is the critical temperature. The so-called retrograde condensation takes place for the negative value
of the aforementioned constant. In Ref.\cite{nak15} the nature of the condensate in external magnetic field and the 
behaviour
of the critical field near the transition temperature were examined. The obtained upturn of the critical
field constitutes the fingerprint of the strong coupling. In that study $\alpha$ has been found 
to be limited to positive values.

The organization of the paper is as follows. In Sec.II we start by studying the s-wave holographic zero temperature 
insulator/superconductor phase transition using the solitonic AdS background. The chemical potential is 
the control parameter of this phase transition. In Sec.III, black hole background 
taken as the gravity configuration allows for the analysis of the transition observed 
for $\mu >\mu_c$ from holographic metal at high temperatures ($T>T_c$) to holographic superconductor 
at temperatures below $T_c$. The effect of the {\it dark matter} sector on the insulator/superconductor
transition of the droplet is studied in Sec.IV. We end the paper with summary and discussion
of the obtained results in the light of variety of the existing metal/insulator and metal/superconductor 
transitions  in condensed matter systems. We limit our analysis of the phase transitions to the probe limit.

\section{Phase transition insulator/superconductor}
In this section we analyze the model of s-wave holographic insulator/superconductor phase transition 
in five-dimensional spacetime, where the matter sector is coupled to another $U(1)$-gauge field, 
representing the {\it dark matter} sector. In the probe limit, we setup the considered model  
in the AdS soliton background \cite{hor98}, which line element is subject to the relation
\be
ds^2 = - r^2~dt^2 + L^2~{dr^2 \over f(r)} + f(r)~d\varphi^2  + r^2 ~(dx^2 + dy^2),
\ee
where  $f(r) = r^2 - r_0^4/r^2$. The geometry resembles a cigar, if one gets rid of
$(r,~\varphi)$-coordinates, with a tip located at $r=r_0$.
The AdS soliton solution is achieved by making two Wick rotations on a five-dimensional
AdS Schwarzschild black hole line element. The asymptotic AdS spacetime tends to $R^{1,2} \times S^1$
topology near the boundary. A conical singularity at $r_0$ can be removed by the Scherk-Schwarz
transformation of $\varphi$-coordinate, i.e., $\varphi \sim \varphi + {\pi~L/r_0}$.
Due to the compactification of $\varphi$-direction, the AdS solitonic background alllows for a 
description of a three-dimensional field theory with a mass gap, which echoes an insulator
in the condensed matter physics. The temperature in the solitonic background is equal to zero.

Without loss of generality we put $L = 1$ and for simplicity we assume that  $A_{t} = \phi(r),~B_{t} = \eta(r)$ 
and $\psi = \psi(r)$. The underlying system of differential equations for scalar and gauge fields yields
\ben
\p_r^2 \psi &+& \bigg( {\p_r f \over f} + {3 \over r} \bigg)~\p_r \psi + \bigg(
{q^2~\phi^2 \over r^2~f} - {m^2 \over f} \bigg)~\psi = 0, \\
\p_r^2 \phi &+& \bigg( {\p_r f \over f} + {1\over r} \bigg)~\p_r \phi - 
{2~q^2~\psi^2~\phi \over {\tilde{\alpha}}~f} = 0,\\
\p_r \eta &=& {c_1 \over r~f} - {\alpha \over 2}~\p_r \phi,
\een
where we set $\tilde{\alpha} = 1 - \frac{\alpha^2}{4}$ and  $c_{1}$ as an integration constant.

Next we impose the boundary conditions on the adequate quantities.
Namely, at the tip of the AdS soliton we demand that the solutions will be provided by
\ben \label{nbc}
\psi &=& \psi_0 + \psi_1 (r-r_0) + \psi_2 (r - r_0)^2 + \dots, \\ 
\label{nbc1}
\phi &=& \phi_0 + \phi_1 (r-r_0) + \phi_2 (r - r_0)^2 + \dots,
\een
where $\psi_m$ and $\phi_m$, for the range $m = 0,~1,~2,\dots$, are constants. Moreover,
in order to achieve the finiteness of the considered quantities, one has to fulfill the Neumann-like 
boundary conditions  ($\psi_1=0$ and $\phi_1=0$). Contrary to the AdS-black hole case, 
where at the event horizon $\phi$ is equal to zero, here it can acquire a non-zero value at the tip of the AdS soliton.
On the other hand, near $r \rightarrow \infty$, we have the following behaviours:
\be
\psi = {\psi^{-} \over r^{\la_{-}}} + {\psi^{+} \over r^{\la_{+}}},
\qquad
\phi = \mu - {\rho \over r^2},
\label{bc}
\ee
where $\mu$ and $\rho$ stand for the chemical potential and charge density in the dual theory,
while $\la_{\pm} = 2 \pm \sqrt{4 +m^2}$. The coefficients $\psi^{\pm}$ are responsible for the vacuum
expectation values of the operators $<\cO_{\pm}>$ dual to the scalar field. One can impose the conditions that
either $\psi^{-}$ or $\psi^{+}$ vanishes \cite{sio10}. In what follows we shall assume that
$\psi^{-}$ vanishes and consider $\psi^{+}=<\cO_i>$ with $<\cO_i>$ denoting the expectation value 
of the corresponding CFT operator. 

It will be convenient to rewrite the above equations in terms of $z = r_{0}/{r}$ variable. They 
reduce to the forms
\ben
\psi'' &+& \bigg( {f' \over f} - {1 \over z} \bigg)~\psi' + \bigg(
{q^2~\phi^2 \over z^2~f} - {m^2~r_0^2 \over z^4~f} \bigg)~\psi = 0, \\
\label{fi10}
\phi'' &+& \bigg( {f' \over f} + {1\over z} \bigg)~\phi' - 
{2~q^2~\psi^2~\phi~r_0^2 \over {\tilde{\alpha}}~f~z^4} = 0,\\
\eta' &=& - {c_1 \over z~f} - {\alpha \over 2}~\phi',
\een
where the prime denotes the derivation with respect to $z$-coordinate.

\subsection{Critical chemical potential}
It was revealed in Ref.\cite{nis10} that when the chemical potential exceeds a critical value, the condensation 
will set in. This state can be interpreted as a superconductor phase. In the case when $\mu < \mu_c$ the scalar field 
$\psi$ achieves  value close to zero and the phase can be interpreted as the insulator. The system 
has a mass gap, which is connected with the confinement in $(2+1)$-dimensional gauge theory
via performing the Scherk-Schwarz compactification. In the light of these facts, the critical value of 
the chemical potential is the turning point in a superconductor phase transition.

For the chemical potential $\mu=\mu_c$, the scalar field is very small
$\psi \sim 0$ and the  equation (\ref{fi10}) for  the gauge field $\phi$  reduces to the form
\be
\phi'' + \bigg( {f' \over f} + {1\over z} \bigg)~\phi' \sim 0.
\ee
The general solution of this equation can be easily found to read
\be
\phi(z)=d_1+d_2\log\frac{1-z^2}{1+z^2},
\ee  
where $d_1$ and $d_2$ are integration constants.

In order to fulfill the assumed boundary conditions (\ref{nbc1}) at the tip $z=1$, we require
$d_2=0$. Thus $\phi $ has the constant value $\mu$, when $\psi(z) = 0$. Moreover, from the equation (\ref{bc}),
one obtains that in the considered case $\rho = 0$. These results are in accord with the numerical 
analysis presented in \cite{nis10}.

\par
By virtue of the above, as $\mu \rightarrow \mu_c$, we have
\be
\psi'' + \bigg( {f' \over f} - {1 \over z} \bigg)~\psi' + \bigg(
{q^2~\mu^2 \over z^2~f} - {m^2~r_0^2 \over z^4~f} \bigg)~\psi = 0.
\ee
By introducing a trial function \cite{sio10} near the boundary $z=0$, in the form 
$\psi(z) = <\cO_i>z^{\la_i}F(z)$, where $i = +$ or $-$, and by imposing the boundary conditions
$F(0)=1$ and $F'(0)=0$, the underlying equation can be brought to the following form
\be
( p(z)~F'(z))' - q(z)~F(z) + \mu^2~r(z)~F(z) = 0,
\ee
where the various terms in the above relation are provided by
\ben
p(z) &=& z^{2 \la_i -1}~f,\\
q(z) &=& - z^{2 \la_i -2}~\bigg( {\la_i(\la_i -1)~f \over z} + 
\bigg( {f' \over f} - {1 \over z} \bigg)~\la_i~f - {m^2~r_0^2 \over z^3}\bigg),\\
r(z) &=& q^2~ z^{2 \la_i -3}.
\een
According to the Sturm-Liouville eigenvalue problem, we can specify $\mu^2$ as a spectral parameter and  estimate 
its minimum eigenvalue by varying the following functional
\be
\mu^2 = {\int_{0}^{1}dz~[ F'(z)^2 ~p(z) + q(z)~F^2(z)] \over \int_{0}^{1}dz~r(z)~F^2(z)}.
\label{funct-mu2}
\ee
The trial function will be set in the form $F(z) = 1 - az^2$. Importantly, the  critical value 
of the chemical potential is unaffected by the {\it dark matter} sector parameters.
The value of the $\mu^2$ in equation (\ref{funct-mu2}) depends on the parameter $a$
entering the trial function. Changing $a$ we  find numerically minimal value of $\mu^2$ for  $a=a_{min}$.
Both $a_{min}$ and $\mu_c^2=\mu^2(a_{min})$ depend on the parameters of the model - in particular $m^2$.
In Table \ref{tab1} the critical chemical potential  has been  presented for a  few typical values of $m^2$
fulfilling the Breitenlohner-Freedman bound $m^2>-d^2/4=-16/4$ required for the stability of the AdS$_{d+1}$ spacetime.
The same result was analytically obtained  in s-wave holographic/superconductor phase transition studies 
in Einstein-Maxwell scalar theory \cite{cai11c}, and is in accord with the numerical 
examinations provided in Ref.\cite{nis10}.
On the other hand, in Gauss-Bonnet gravity, for the transition in question, one observes
the critical potential increase with the growth of curvature corrections, for the same mass of scalar field.
For the fixed value of the strength of curvature corrections, with the increase of scalar field mass, 
the critical potential becomes larger \cite{pan11}.
\begin{table}[h!]
\begin{center}
\begin{tabular}{|c|c|c|c|c|}
  \hline
  $\lambda_+$           & $ m^2 $               &  $ \mu_c$   &$a_{min}$\\
  \hline
  $~~ \frac{5}{2}~~ $ & $~~ -\frac{15}{4}~~$  &  $~~ 1.890~~ $ &0.330 \\
  \hline 
  $~~ \frac{6}{2}~~ $ & $~~ - \frac{12}{4}~~$ &  $~~ 2.398 ~~$ &0.371\\
  \hline
  $~~ \frac{7}{2}~~ $ & $~~ - \frac{7}{4} ~~$ &  $~~ 2.903~~$ &0.407\\
 \hline
  $~~ \frac{8}{2}~~ $ & $~~  0 ~~$ &  $~~ 3.406~~$ &0.439\\
 \hline
\end{tabular}
\end{center}
\caption{Values of the critical chemical potential together with $a$ parameters
minimizing the functional (\ref{funct-mu2}). In all the above examples we put $q=1.0$ and $r_0=1$. 
The numerical values of $m^2$ and $\lambda_+$ are chosen for illustration purposes.
They obey the physical requirement that the masses $m^2$ do fulfill the
Breitenlohner-Freedman bound $m^2>-d^2/4=-16/4$ required for the stability of the AdS$_{d+1}$ spacetime.}
\label{tab1}
\end{table}

\subsection{Critical phenomena}

In this subsection, we shall concentrate on studies of the critical exponent for condensation operator as well 
as on the mutual relations between the charge density $\rho$  and the chemical potential.
The question we are asking here is how the order parameter of the superconductor $i.e.$ $<\cO_i>$ and the charge density $\rho$  
 depend on the distance ($\mu-\mu_c$). 
Having in mind the form of the scalar field ($\psi(z) = <\cO_i>z^{\la_i}F(z)$, where $F(z)$ is the trial function introduced 
earlier)  near the boundary $z=0$,  when $\mu \rightarrow \mu_c$, the  relation for gauge $A_t = \phi(r)$ field can be rewritten as 
\be
\phi'' + \bigg( {f' \over f} + {1\over z} \bigg)~\phi' - 
{2~q^2~r_0^2 \over {\tilde{\alpha}}~f} <\cO_i>^2~{z^{2\la_i -4}}~F^2(z)~\phi = 0.
\label{agauge}
\ee

To proceed further, let us remind that for $\mu$ slightly above the critical
value the condensation scalar operator $<\cO_i>$ is very small. This enables us to seek the solution
in the form 
\be
\phi(z) \sim \mu_c ~+~ <\cO_i>~\chi(z) + \dots
\label{fiz}
\ee
In order to recover the previous result $\phi(z)=\mu$, we have to impose boundary condition $\chi(1)=0$.
On the other hand, close to the boundary $z=0$ one expands the function 
$\chi(z)=\chi(0)+\chi'(0)z+\frac{1}{2}\chi''(0)z^2+\ldots$, rewrites the relation (\ref{bc}) as
\be
\phi(z) \simeq \mu - \rho~z^2 \simeq
\mu_c + <\cO_i>\bigg( \chi(0) + \chi'(0)z + {1 \over 2}\chi''(0)~z^2 + \dots \bigg),
\label{expan}
\ee
Comparing the coefficients of the $z^0$ and $z^1$-terms, in the above equation, one  obtains the relations 
\ben
&&\mu - \mu_c \simeq~ <\cO_i> \chi(0) \\
&&\chi'(0)=0.
\label{scaloper}
\een

Inserting the relation (\ref{fiz}) into (\ref{agauge}), one can easily find that $\chi(z)$ will satisfy the following
equation: 
\be
\chi''(z) + \bigg( {f' \over f} + {1\over z} \bigg)~\chi'(z) - 
{2q^2~r_0^2 \over {\tilde{\alpha}}~f} ~{z^{2\la_i -4}}~F^2(z)\mu_c<\cO_i>
- {2q^2~r_0^2 \over {\tilde{\alpha}}~f} ~{z^{2\la_i -4}}~F^2(z)<\cO_i>^2\chi(z) + \dots  = 0.
\ee
Close to $\mu_c$ the term quadratic in  $<\cO_i>$ is much smaller than the linear one and
may be safely neglected leading to
\be
\chi''(z) + \bigg( {f' \over f} + {1\over z} \bigg)~\chi'(z) = 
{2q^2~r_0^2 \over {\tilde{\alpha}}~f} ~{z^{2\la_i -4}}~F^2(z)\mu_c<\cO_i>.
\label{ga2}
\ee
In the next step, let us redefine $\chi(z)$ function by the new one $\xi(z)$ multiplied by the adequate factor
\be
\chi(z) = 2~{<\cO_i>~\mu_c \over {\tilde{\alpha}}} ~\xi(z).
\ee
It remains to be checked if the new definition of $\chi(z)$ enables us to get rid of the ${\tilde{\alpha}}$
in the equation (\ref{ga2}) and to extract the quantity $<\cO_i>$.
It can be inspected that $\xi(z)$ will satisfy the following relation 
\be
\xi'' + \bigg( {f' \over f} + {1\over z} \bigg)~\xi' 
- {q^2~r_0^2 \over f} ~{z^{2\la_i -4}}~F^2(z) = 0,
\label{cc1}
\ee
and consequently the scalar operators $<\cO_i>$ imply
\be
<\cO_i> = \sqrt{{(\mu - \mu_c)~{\tilde{\alpha}} \over 2~\mu_c~\xi(0)}}.
\label{scalop}
\ee
In order to find $\xi(0)$  it will be helpful to rewrite the equation (\ref{cc1}) in the form
\be
\bigg(f~z~\xi' \bigg)' = {q^2~r_0^2} ~{z^{2\la_i -3}}~F^2(z).
\ee
Then, having in mind the fact that $\xi'(1)=0$, leads to the conclusion that
$\xi(0)$ is provided by 
\be
\xi(0)= c_1 - \int_0^1 {dz \over f~z}~\bigg(c_2 + \int_1^z dy ~
q^2~r_0^2~y^{2\la_i -3}~F^2(y) \bigg).
\label{xi0}
\ee
The integration constant $c_1,~c_2$ are determined by the boundary conditions imposed on $\chi(z)$-function.
We relegate their determination to the Appendix A.

On the other hand, the above relations reveal that the operators $<\cO_i>$ yield
\be
<\cO_i> \simeq ~\Gamma~(\mu - \mu_c)^{1 \over 2},
\ee
where the $\Gamma$ factor contains information of the dependence on {\it dark matter} sector.
The bigger $\tilde{\alpha}$ (the smaller value of the $\alpha$-coupling constant we take)
we consider, the greater factor one obtains. 

\par
Moreover, our analytical results show that the holographic s-wave insulator/superconductor
phase transition represents the second order phase transition, with the critical exponent of the system
attaining the mean-field value $1/2$.
The same conclusions were achieved in the case of
of the ordinary s-wave holographic insulator/superconductor phase transition studies in Refs.\cite{cai11b,nis10}.
On the other hand, the same form of the dependence was also obtained in Gauss-Bonnet theory \cite{pan11}, confirming
the previous numerical results \cite{pan10,liu10}. The Gauss-Bonnet coupling constant connected with the
influence of curvature corrections, does enter in the multiplier factor of the scalar operators, but
the critical value of the exponent takes the mean-field value.

\begin{figure}[h]
\includegraphics[scale=0.5]{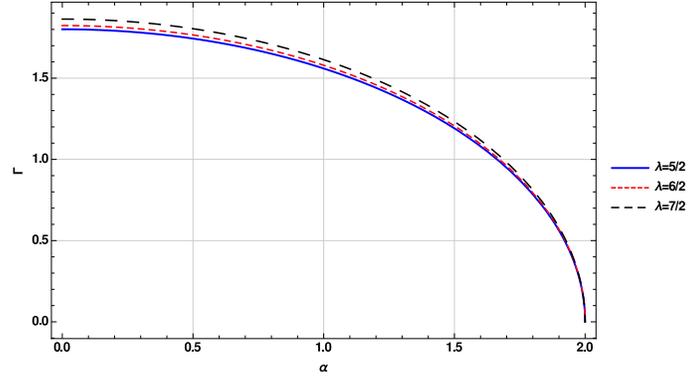}
\caption{(color online). The dependence of the prefactor of the condensate function on the coupling constant $\alpha$. 
The trial function was chosen as $F = 1 - az^2$
and the charge was set equal to $q=1$. }
\label{fig1}
\end{figure}

Next, we  find the dependence of the charge density $\rho$ on the critical chemical potential.
In order to calculate $\rho$, we use (\ref{scaloper}) which implies 
$\xi'(0)=0$ together with the previous requirement $\xi(1)=0$ being subject to the boundary condition.
Comparison of the adequate coefficients of $z^2$-order in equation (\ref{expan}), gives the 
relation for  the charge density in the following form 
\be
\rho = - {<\cO_i> \over 2}~\chi''(0).
\label{denc}
\ee
In order to find $\chi''(0)$
we rewrite equation (\ref{ga2}) in the form which implies
\be
\bigg( f~z~\chi' \bigg)' - {2q^2~r_0^2 \over {\tilde{\alpha}}} ~{z^{2\la_i -3}}~F^2(z)\mu_c<\cO_i> = 0.
\ee
Integrating both sides of it and taking into account the aforementioned boundary conditions, one obtains
\be
\chi''(0) =
{\chi'(z) \over z} \mid_{z \rightarrow 0} = - 2<\cO_i>~{q^2~\mu_c \over \tilde{\alpha}}
~\int_0^1 dz~z^{2\la_i - 3}~F^2(z).
\label{li}
\ee
Then, by virtue of the equations (\ref{li}) and (\ref{denc}), having in mind the relation (\ref{scalop}), one arrives at 
\be
\rho = (\mu - \mu_c)~{\tilde{ B}},
\ee
where the quantity $\tilde{ B}$ yields
\be
\tilde{ B} = {q^2 \over 2~\xi(0)}~\int_0^1 dz~z^{2\la_i -3}~F^2(z).
\ee 
In view of the above relations, the charge density is proportional to the difference $\Gamma_1~(\mu -\mu_c)$, where
the factor is independent on $\tilde{\alpha}$ characterizing {\it dark matter} sector and accomplishes the ordinary dependence 
of the form $\rho \sim (\mu-\mu_c)$ achieved analytically in \cite{cai11b} and numerically in Refs. \cite{hor10,akh11}. 
Some typical values of $\tilde{B}$ factor are presented in Table \ref{tab2}. The dependence of the $\Gamma$ factor 
on the {\it dark matter} sector coupling $\alpha$ is depicted in Fig.\ref{fig1}.

In the case of Gauss-Bonnet gravity sector, the factor standing in front of $(\mu-\mu_c)$,
is function of mass and Gauss-Bonnet coupling constant. However, the form of the linear dependence survives
\cite{pan11,pan10}. 

\begin{table}[h!]
\begin{center}
\begin{tabular}{|c|c|c|c|c|c|}
  \hline
  $\lambda_+$           & $ m^2 $                & $ \mu_c$       & $\xi(0)$      & $\frac{\Gamma}{\sqrt{\tilde{\alpha}} } $     & $\tilde{B}$  \\
  \hline
  $~~ \frac{5}{2}~~ $ & $~~ -\frac{15}{4}~~ $  &  $~~ 1.890~~$ & $~~0.081~~$  & $~~1.801~~$                                 & $~~1.329~~$ \\
  \hline 
  $~~ \frac{6}{2}~~ $ & $~~ - \frac{12}{4}~~ $ &  $~~ 2.398~~$ & $~~0.062~~$  & $~~1.823~~$                                 & $~~1.144~~$ \\
  \hline
  $~~ \frac{7}{2}~~ $ & $~~ - \frac{7}{4}~~ $  &  $~~ 2.903~~$ & $~~0.049~~$  & $~~1.863~~$                                 & $~~1.029~~$ \\
  \hline
  $~~ \frac{8}{2} ~~$ & $~~ 0 ~~$              &  $~~ 3.406 ~~$& $~~0.040 ~~$ & $~~ 1.913 ~~$                                & $~~ 0.948~~$\\
  \hline
\end{tabular}
\end{center}
\caption{Values of the prefactors for the condensate and charge density. 
In all above examples $q=1.0$ and trial function was of the form $F = 1 - az^2$.}
\label{tab2}
\end{table}

\section{Holographic metal/superconductor phase transition}
In this section we shall scrutinize the problem of holographic s-wave metal/superconductor phase transition
at low temperatures provided by the black hole background.
The problem of s-wave holographic superconductor with {\it dark matter} sector in the context of the
backreaction of matter fields on the gravitational background was investigated in \cite{nak14}, where the 
critical temperature was found. On the other hand, for n-dimensional gravitational background, 
the expectational value of the scalar operator and the influence of magnetic field on the holographic
superconductor was analyzed \cite{nak15}.
For the completeness of the investigations, using quite different methods, we elaborate
the dependence of the critical temperature and the scalar operator
on the presence of the {\it dark matter} sector.

To commence with, one considers the background of five-dimensional black hole given by the line element
\be
ds^2 = - g(r)~dt^2 + {dr^2 \over g(r)} + {r^2 \over L^2}~(dx^2 + dy^2 + dz^2),
\ee
where $g(r) = r^2/L^2 - r_+^4/r^2 L^2$. The Hawking temperature for the black hole 
has the form $T_{BH} = r_+/\pi$. In $z$-coordinate the equations of motion imply
\ben
\label{eqIIIpsi}
\psi'' &+& \bigg( {g' \over g} - {1 \over z} \bigg)~\psi' + \bigg(
{q^2~\phi^2 \over g^2} - {m^2 \over g} \bigg)~{r_+^2 \over z^4}~\psi = 0, \\
\phi'' &-& {1\over z}~\phi' - 
{2~q^2~\psi^2~\phi~r_+^2 \over {\tilde{\alpha}}~g~z^4} = 0,\\
\eta' &=& - {d_1~z \over r_+^2} - {\alpha \over 2}~\phi',
\een
where the prime denotes the derivation with respect to $z$-coordinate.
In the following, as in the preceding sections, we set $L=1$. 
In order to solve the above equations we need to impose the adequate boundary conditions. At the black hole horizon $z=1$,
it is required that $\phi(1)=0$ and $\psi(1)$ should be finite. The first requirement is needed for the 
$U(1)$-gauge field to have the finite form,  the second one exhibits that the black hole has a scalar hair on the event horizon.

When the temperature $T$ tends to the critical value $T_c$ from below, the condensation approaches zero, $\psi \rightarrow 0$.
In this limit we write the equation for $\phi$ field as
\be
\phi'' - {1\over z}~\phi' \simeq 0.
\ee
Its general solution is of the form $\phi(z)=c_1+c_2z^2$, which together with the aforementioned boundary condition 
at the horizon leads to $\phi(z)=c_1(1-z^2)$ .
Then, having in mind that near the boundary of the bulk the fields behave as
\ben
\phi &\rightarrow& \mu - {\rho \over r^2}=\mu-\rho z^2/r_+^2,\\
\psi &\rightarrow& {\psi^{-} \over r^{\Delta_{-}}} + {\psi^{+} \over r^{\Delta_+}},
\een
one arrives at the conclusion that near the critical temperature the $A_t$ gauge field component
will behave as $\phi \simeq \la~r_+~(1-z^2)$, where we have denoted by $\la = \rho/r_+^3$.
$\mu$ and $\rho$ have the same interpretation as in the preceding sections, i.e., denote
the chemical potential and the charge density, respectively. Moreover, $\Delta_{\pm}$ 
have the same values as in Sec.II, but in order to distinguish 
the nature of the phase transition, we  set the new notation for the old quantity.
In what follows we concentrate on the $\Delta=\Delta_+$ and the masses $m^2$ 
fulfilling the Breitenlohner-Freedman  spacetime stability condition.
It is important to note that we are looking for the parameter $\la$ which defines the 
value of both $\mu$ and $\rho$. It depends on the temperature given by $T=T_{BH}=r_+/\pi$.

On the other hand, the scalar field $\psi$ can be cast near the boundary in the form 
\be
\psi \mid_{z \rightarrow 0} \simeq~ <\cC>~{z^\Delta \over r_+^\Delta}~G(z),
\ee
where, we set $G(0)=1,~G'(0)=0$. 
Inserting this expression into the equation (\ref{eqIIIpsi}) and using $\phi = \la~r_+~(1-z^2)$
one finds the scalar field equation of motion, which can be easily rewritten as the Sturm-Liouvile eigenvalue problem
\be
( \cP(z)~G'(z))' - \cQ(z)~G(z) + \la^2~\cR(z)~G(z) = 0,
\ee 
where one has defined the quantities
\ben
\cP &=& z^{2 \Delta -1}~g,\\
\cQ &=& - \Delta (\Delta -1)~g~z^{2 \Delta - 3} - \bigg(
{g' \over g} - {1 \over z} \bigg)~\Delta~z^{2 \Delta - 2}~g + m^2~r_+^2~z^{2\Delta -5},\\
\cR &=& {z^{2\Delta -5} \over g}~r_+^4~(1-z^2)^2~q^2.
\een 
The minimum value of $\la^2$ can be found from variation of the functional given by
\be
\la^2 = {\int_{0}^{1}dz~[ G'(z)^2 ~\cP(z) + \cQ(z)~G^2(z)] \over \int_{0}^{1}dz~\cR(z)~G^2(z)}.
\label{funct-la2}
\ee
where the trial function is assumed to be given by $G(z) = 1 - a~z^2$.
Note that in the variable $z$, the function $g(z)=r_+^2(1/z^2-z^2)$ and the value of event horizon radius $r_+$
factors out of the expression for $\la^2$ making it independent on $T_{BH}$.
Because of that and the fact that the analysis is valid close to the transition temperature $T_c \sim T_{BH}$, one readily finds that
\be                
T_c = \rho^{1/3}~\bigg( {1 \over \pi^3~\la_{min}}\bigg)^{1/3}.
\ee
It is important to note that the value $\la_{min}$ results from the variation of Sturm-Liouville functional (\ref{funct-la2}).
Holographic superconductors with different values of $\rho$ are characterized by different transition temperatures. 
This reminds the charge carrier concentration dependence of the local pairing superconductors \cite{micnas1990}.
However, the charge density dependence of the critical temperature $T_c\propto \rho^{1/3}$ found here markedly differs from the 
known dependence $T_c\propto n^{2/3}$ for the Bose-Einstein condensation of the low density $n$ superconductors 
with local  pairs of charged hard core bosons \cite{micnas1990}. Whether the difference is the hallmark of the strong coupling 
behavior remains to be seen. In fact it has been suggested \cite{har08b} that the condensation transition in
holographic models is closer to the Bose-Einstein condensation than BCS-like  symmetry breaking phase transition.  
It has to be reminded that the critical temperature does not depend on the carrier concentration $n$ 
in the standard weak coupling superconductors described by the BCS theory. Instead it depends on the density of states
at the Fermi level \cite{BCS}.

It can be observed that the critical temperature does not depend on the  {\it dark
matter} sector. This conclusion is in accord with our previous studies \cite{nak14}, where it was revealed that
the backreaction effects introduce the dependence of $T_c$ on the {\it dark matter} sector.
In the case under consideration we restrict our investigations to the probe limit case, therefore
no influence is spotted. 

In the Table {\ref{tab3}} we have presented the results of the calculation of the superconducting transition 
temperature assuming the charge density $\rho=1$. The presented theory is valid for arbitrary allowed
values of $m^2$, but for illustration we have chosen some  exemplary values of it and calculated $\la^2$ and the critical temperature $T_c$. 
Both of these parameters are presented in the Table {\ref{tab3}} together with the value of the 
parameter $a$, which minimizes the functional (\ref{funct-la2}).

\subsection{Condensation values}
In this subsection our main task will be to find the influence of the {\it dark matter} sector field on
the condensation operator. Near the critical temperature, the equation for gauge field can be rewritten as 
\be
\phi'' - {\phi' \over z} = {2~q^2~r_+^2~\phi \over \tilde{\alpha}~g}~z^{2\Delta -4}~G^2(z)~\cA,
\ee
where we have denoted $\cA = <\cC>^2/ r_+^{2\Delta}$.
Having in mind the fact that near the critical temperature the $\cA$ quantity is small, one expands
$\phi$ near $z \rightarrow 0$. On this account, we can write the following: 
\be
{\phi \over r_+} = \la~(1-z^2) + \cA~\chi(z) + \dots.
\label{pp}
\ee
In the next step, comparing the coefficients in $z^2$-order terms, we reveal that
\be
{\rho \over r_+^3} = \la - {\cA \over 2}~\chi''(0).
\ee 
On the other hand, considering the relation (\ref{pp}) and the equation of motion for $\phi$ field, we get
\be
\chi'' - {\chi' \over z} = {2~\la~q^2~r_+^2 \over \tilde{\alpha}~f}~(1-z^2)~z^{2\Delta-4}~G^2(z).
\ee
Consequently, proceeding as in the last section, it can be verified that the following is satisfied
\be
\chi''(0) = {\chi'(z) \over z} \mid_{z \rightarrow 0} =
- 2~\la~\int_0^1 dz~{q^2 ~r_+^2 \over \tilde{\alpha}~f}~(1-z^2)~z^{2\Delta-5}~G^2(z).
\ee
On evaluating the expression for $\cA$, in the case when $T \rightarrow T_c$, the condensation
operator in question is provided by
\be
<\cC> = \sqrt{2~\tilde{\alpha} \over \cB}~(\pi~T_c)^\Delta~\sqrt{1 - {T \over T_c}}
= \sqrt{\tilde{\alpha}}~<\cC>_{no~ dark~ sect},
\ee
where by $<\cC>_{no~ dark~ sect}$ we denoted the value in the theory without {\it dark matter} sector. The term 
$\cB$ is given by the relation 
\be
\cB = 2\int_0^1 dz~{q^2~r_+^2 \over g}~(1-z^2)~z^{2\Delta-5}~G^2(z).
\ee
One can see that the condensation operator depends on the $\alpha$ constant coupling of the
{\it dark matter} sector. The bigger is the $\alpha$-coupling, the easier condensation forms. 
From the point of view of the AdS/CFT correspondence, the operator in question can be interpreted as
the operator for pairing mechanism. The bigger expectation value it achieves, the harder condensation occurs.
In order to better understand the dependence of the condensate on the dark matter sector we factor 
out its dependence on the rest of the parameters in the following way:
\ben
< \mathcal{C} >_{norm} = \frac{<\mathcal{C}>}{T_c^{\Delta}} 
\equiv \sqrt{\tilde{\alpha}} \tilde{C} \sqrt{1 - \frac{T}{T_c}}, 
\label{prefC}
\een 
where $\tilde{C} \equiv \sqrt{\frac{2}{\mathcal{B}}} \pi^{\Delta}$ 
and $<C>_{norm}$ represent renormalized value of the condensate.
The factor $\tilde{C}$  depends on all the remaining parameters (except temperature) and its 
typical values are presented in Table \ref{tab3}.
These facts can potentially constitute the way of determining the {\it dark matter} sector 
in future 'possible' superconductor experiments.


We remark that, in the case of Gauss-Bonnet theory \cite{li11}, the value of $<\cC>$ is dependent
on the higher curvature term corrections. When it grows, the value of the operator also increases. This conclusion 
is in agreement with the previous studies (see, e.g., \cite{pan10} and references therein).

\begin{table}[h!]
\begin{center}
\begin{tabular}{|c|c|c|c|c|c|}
  \hline
  $~~\Delta~~$       & $ m^2 $              &$\lambda^2$ & $a_{min}$ & $ T_{c}(\rho = 1)$   & $\tilde{C} $  \\
  \hline
  $~~ \frac{5}{2} ~~$ & $~~ -\frac{-15}{4} ~~$ &    ~9.586~ & ~0.619~     &  $ 0.218$          & $~~55.67~~$   \\
	\hline
  $~~ \frac{6}{2} ~~$ & $~~ -\frac{12}{4} ~~$ &    ~18,22~ & ~0.721~     &  $ 0.196$          & $~~137.8~~$   \\
	\hline
  $~~ \frac{7}{2} ~~$ & $~~ -\frac{7}{4} ~~$ &    ~30.50~ & ~0.797~     &  $ 0.180$          & $~~326.7~~$   \\
  \hline 
  $~~ \frac{8}{2} ~~$ & $~~ 0 ~~$            &    ~46.89~ & ~0.853~     &  $ 0.168$          & $~~748.3~~$  \\
  \hline
\end{tabular}
\end{center}
\caption{Calculated values of the prefactors of $\la^2$ minimizing the functional (\ref{funct-la2}), the superconducting transition temperature
of the superconductor with charge denisity $\rho=1$ and the prefactor $\tilde{C}$ defined in Eq. (\ref{prefC}). 
In all the above examples we assumed $q=1.0$ and selected the trial function  as $G = 1 - az^2$. We also provide the 
parameter $a$ minimizing the functional (\ref{funct-la2}).}
\label{tab3}
\end{table}

\section{Holographic droplet in s-wave insulator/superconductor phase transition}
The term superconducting droplet  refers to the solutions that are confined in space and rapidly decay
at large distances. This happens if the studied superconductor is exposed to strong external magnetic field.
The size of the confining region diminishes with the increase of the magnetic field. 
In this section we shall investigate the onset of the transition by studying the marginally stable modes of scalar perturbations.
It has been shown earlier \cite{cai11a,cai11b,cai14} that  in the holographic approach the marginally stable 
modes signal the appearance of the insulator-superconductor transition.

Here we are interested in the insulator - superconductor phase transition in the AdS solitonic background coupled 
to the {\it dark matter} sector.
The presence of the  magnetic field introduced $via$ dark matter potential enables us to examine 
the droplet solution via the aforementioned technique.

\par
It turned out that the quasi-normal modes (QNMs) technique occurred as a method of examining stability of a spacetime
background \cite{fro98}. In the case when the imaginary part of QNMs is negative, the modes decrease in time and result in
disappearance of perturbations (background is stable against perturbations). On the other hand,
when imaginary part is positive, the background is unstable against the perturbations in question. 
The marginally stable modes are the modes which frequencies go to zero ($\omega = 0$) near the critical point
of the phase transition. Their appearance thus signals the phase transitions \cite{gub08,gub08b}.

\par
To commence with, we consider AdS soliton metric, taking into account symmetry of the problem in question
and rewrite the line element used in Sec.II in the following coordinates $(t,~,r,~\tilde{\rho},~\varphi,~\theta)$ as   
\be
ds^2 = - r^2~dt^2 + {dr^2 \over f(r)} + f(r)~d\varphi^2  + r^2 ~(d\tilde {\rho}^2 + \tilde{\rho}^2 ~d\theta^2),
\ee
Further, we assume the existence, in addition to a constant chemical potential bounded with $A_t$ gauge field component,
the $B_\theta$ potential which corresponds to the {\it dark matter} gauge field and is proportional 
to the constant value of magnetic field $B$ 
\be
A_t = \mu, \qquad B_\theta = {1 \over 2}B~\tr^2.
\ee
The above ansatz stems from the fact that one considers
gauge sector close to the critical point of the phase transition, i.e.,
$\mu \sim \mu_c$ and $\psi \sim 0$, as well as, the implementation of the polar coordinates in order
to envisage the symmetry of the problem in question \cite{cai11b}.

Having in mind the exact form of $B_\theta$, given by the above relation,
we calculate from equation of motion
\be
\na_\mu B^{\mu \nu} + {\alpha \over 2}~\na_\mu F^{\mu \nu} = 0,
\ee
the $A_{\theta}$ component, which yields
\be
A_{\theta} = {D_1~r~\tr^2 \over \alpha} - {B \over \alpha}~\tr^2 + D_2.
\ee
$D_1$ and $D_2$ are integration constants. To proceed further, we assume 
$A_\theta$ to be the function of $\tr$ only, which implies that $D_1$ and $D_2$ have to be equal zero.
Our studies are devoted to the probe limit, i.e., the $U(1)$-gauge fields and scalar one do not backreact
on the AdS soliton background metric. 
Without the condensate ($i.e.$ for $\psi=0$) the solution of the equations of motion for the 
$A$ gauge field components are given by the $A_\theta=-{B \over \alpha}~\tr^2$ and $A_t=\mu$.
We are interested in finding the solution of $\psi$ equation close to the critical chemical potential $\mu \sim \mu_c$, where the value of the scalar field reaches nearly zero. 
It implies that one
can treat $\psi$ field as a probe into the background consisting of AdS Schwarzschild soliton with constant electric and magnetic field.\\
The equation describing $\psi$ field yields
\be
\na_\mu\na^\mu \psi - q^2~A_\mu A^\mu ~\psi - m^2~\psi = 0.
\ee
By virtue of the above, the explicit form of the equation for $\psi$ field may be written as
\ben
\p^2_r \psi &+& \bigg( {\p_r f \over f} + {3 \over r} \bigg)\p_r \psi + {1 \over f^2} \p^2_{\varphi} \psi
- {1 \over r^2~f} \p^2_t \psi + {1 \over r^2~f} ~{1 \over \tr} \p_{\tr}( \tr~\p_{\tr} \psi) \\ \nonumber
&+&
 {1 \over r^2~f}~\bigg(
q^2~\mu^2 - m^2~r^2 - {q^2~B^2~\tr^2 \over \alpha^2 }\bigg)~\psi = 0.
\een 
In order to solve the above equation we choose an ansatz for $\psi$ field 
\be
\psi = F(r,~t)~H(\varphi)~U(\tr).
\ee
This form enables us to separate variables. After a simple algebra we arrive at the following set of equations 
\ben
\p^2_r F &+& \bigg( {3 \over r} + {\p_r f \over f}\bigg)~\p_r F - {1 \over r^2~f}~\p^2_t F
+
{1 \over r^2~f}~\bigg( q^2~\mu^2 
- m^2~r^2 - {\la^2~r^2 \over f} - k^2 \bigg)~F = 0,\\
{\p^2_{\varphi} H \over H } &=& - \la^2,\\
{1 \over \tr}~\p_{\tr} \bigg( \tr~\p_{\tr} U \bigg) &-& {q^2~B^2~\tr^2 \over \alpha^2}~U = - k^2~U,
\label{eq-U}
\een
From periodicity property $H(\varphi) = H(\varphi + \pi~L/r_0)$ of $H(\varphi)$ we identify  that $\la = 2~r_0~n/L$, 
where $n \in Z$. In what follows without loss of the generality we set $r_0 =1$ and $L=1$ what leads to  $\la = 2n$.
We expect that the lowest mode will be first to condense and result in the most stable solution.

\par
The equation for $U(\tr)$ is a two-dimensional harmonic oscillator one. In order to solve it we recall that 
the function $U(\tr)$ should satisfy the boundary conditions $U(\tr \rightarrow \infty) = 0$. 
It is possible to investigate such kind of differential equation by
Frobenius method. The customary procedure is first to factor out the behaviour of the relevant solutions at infinity
by setting
\be
U(\tilde{\rho})= e^{-\Lambda \tilde{\rho}^2/2}~D(\tilde{\rho}),
\label{urho}
\ee
which results in Hermite'a type of equation. Inserting (\ref{urho}) into the underlying equation, we
obtain the expected type of the differential solution if the condition
\be
\Lambda^2 = {q^2~B^2 \over \alpha^2},
\ee
is satisfied. Then, the resulting equation yields
\be
{1 \over \tr}~\p_{\tr} \bigg( \tr~\p_{\tr} D \bigg) -2~\Lambda~ \tr ~\p_{\tr} D + (k^2 - 2~\Lambda) D = 0.
\ee
In order to find the exact form of $D(\tr)$, one  sets $D(\tr) = \sum_{k=0}^{\infty} a_n~\tr^{k+l}$ (see, e.g., \cite{but73}). 
\par

Note that $D(\tilde{\rho})=const$ is the well know lowest energy solution of the harmonic oscillator. It
leads to the condition 
\be
k^2=2~\Lambda= \mid \frac{2~q~B}{\alpha}\mid.
\ee
 Equation (\ref{urho}) shows that in the presence of magnetic field the superconducting region
is confined in space. For the chosen solution $D(\tr)=const$ it forms a droplet \cite{mil1995} of radius 
\be
<\tr>={ \int d\tr~ \tr~ U(\tr) \over \int d\tr ~U(\tr)} = 
{ 1 \over \sqrt{2\pi~ \Lambda}}= \sqrt{\frac{\alpha}{4~\pi~ q~B}}.
\ee

In the field theory it corresponds to the condensation in  the lowest Landau level \cite{ras1992}.
As far as the time dependence of the $F(r,~t)$ is concerned, we substitute it in the form $F(r,~t) = e^{-i \omega t}~R(r)$. 
The requirement concerning marginally stable modes leads to the condition  $\omega =0$. 
Redefining the coordinates as $z=r_+/r$ enables to arrive at the  equation given by
\be
\p^2_z R(z) + \bigg( {\p_z f \over f} - {1 \over z} \bigg)~\p_z R(z) 
+ {1 \over z^2~f}~\bigg(
q^2~\mu^2 - {2~B~q\over \alpha} - {m^2 \over z^2} - {4~n^2 \over z^2~f} \bigg)~R(z) = 0.
\ee
To solve it close to $\mu_c$ when the field $\psi \approx 0$ we introduce a correction function $\Theta(z)$ in the form
\be
R(z) \mid_{z \rightarrow 0} \sim~ <\cO_i>~z^{\la_i}~\Theta(z),
\ee
with the boundary conditions  $\Theta(0) =1$ and $\Theta'(0) = 0$.
After some algebra, the resulting equation can be 
converted into the standard Sturm-Liouville eigenvalue equation, which can be rewritten as
\be
\p_z\bigg( a(z)~\Theta' \bigg) - b(z)~\Theta + \delta^2~c(z)~\Theta = 0,
\ee
where $\delta^2 = q^2\mu_c^2 - 2~q~B/\alpha$ and the remaining quantities are defined by the relations
\ben
a(z) &=& f~z^{2\la_i-1},\\
b(z) &=& - f~z^{2\la_i-1}~
\bigg( {\la_i(\la_i-1) \over z^2} + \bigg( {\p_z f \over f} - {1 \over z} \bigg)~{\la_i \over z}
- {1 \over z^4~f}~\bigg( m^2 + {4~n^2 \over f} \bigg) \bigg),\\
c(z) &=& z^{2 \la_i -3}.
\een

The eigenvalues of $\delta^2$ can be found by the method of minimizing  the functional 
\be
\delta^2 = q^2~\mu_c^2 - {2~B~q \over \alpha} = {\int_0^1 dz~(\Theta'(z)^2 ~a(z) + b(z)~\Theta(z)^2)
\over \int_0^1 dz~(c(z)~\Theta^2(z))}.
\ee

\begin{figure}[!th]
\centering
\subfloat[$B = 0.1$]{
  \includegraphics[scale=0.4]{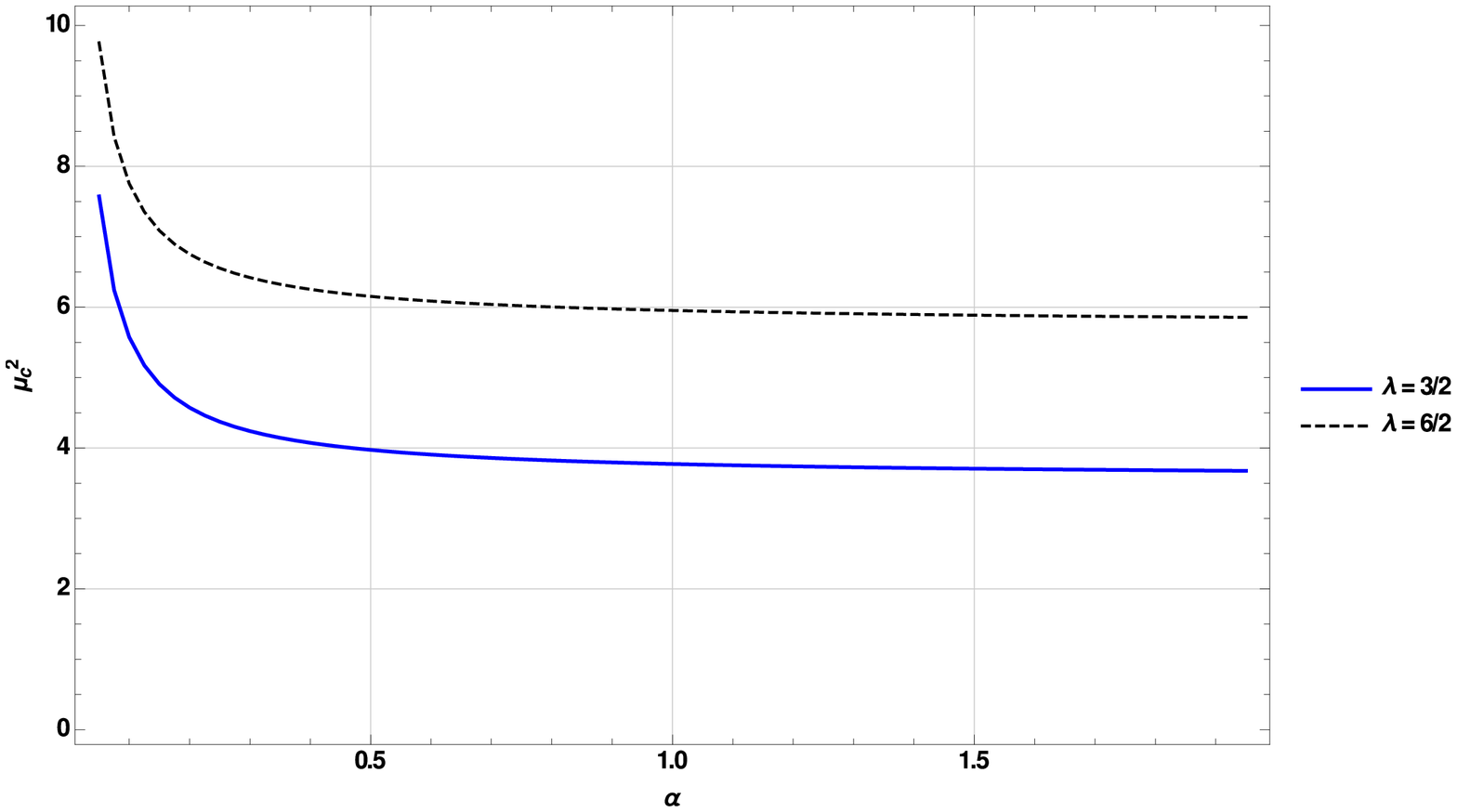}
  }
\subfloat[$B = 1$]{
  \includegraphics[scale=0.4]{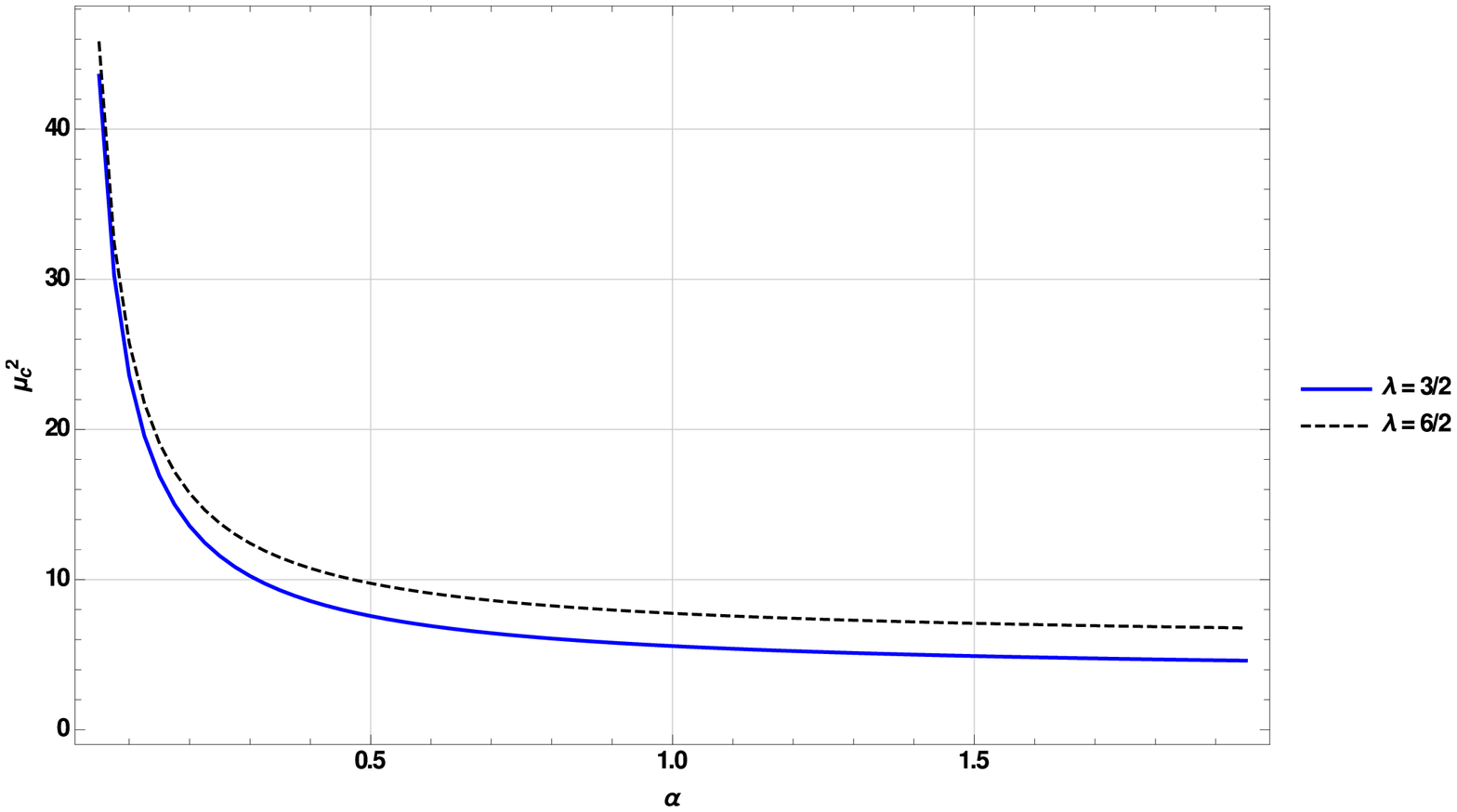}
}
\caption{(color online). The dependence of the critical chemical 
potential on $\alpha$ in the droplet case, for the fixed value of the magnetic field. We set $B=0.1$ (for the left panel) and $B=1$ (for the right panel),
the rest of the parameters are equal to $q=1.0$,~ $n=0$,~ $m^2 = -\frac{15}{4}$($\lambda =  \frac{5}{2}$),~
 $m^2 = -3$~($\lambda = 3$).}
\label{fig2}
\end{figure}

In order to estimate $\delta^2$, we choose function $\Theta(z) = 1- a~z^2.$ 
Minimization of the functional provides an estimation of the value of $\delta$ which depends on $m^2$ and $n$, resulting from the periodicity property of $H(\varphi)$.
The above relation between $\delta^2$, critical chemical potential $\mu_c$ can be rewritten as
\be
\mu_c=\frac{\sqrt{\delta^2+\frac{2qB}{\alpha}}}{q}.
\ee  
It follows that $\mu_c$ depends on  the coupling to the dark matter sector $\alpha$.
Interestingly, for constant magnetic field $B$ the critical chemical potential diverges
for $\alpha\rightarrow 0$. It means that at the constant magnetic field the condensation is harder to occur
for smaller values of $\alpha$. However, one has to remember that the zero value of $\alpha$ is not allowed, as 
simultaneously one has to take $B=0$, so without dark matter field the standard relation, valid
for zero magnetic field \cite{roy13} $\delta^2=q^2\mu_c^2$, is recovered. The increase of the magnetic field
causes the increase of $\mu_c$, which in turn eventuates in the harder condensation.
The aforementioned behaviour is depicted in Fig.\ref{fig2}  for two values of the magnetic field. The
discussed increase of $\mu_c$ for small $\alpha$ is clearly visible both for $B=0.1$ and for $B=1$.

\section{Discussion and Conclusion}
The main aim of our paper is to find the quantitative or at least qualitative  imprints of the {\it dark matter} sector 
on the properties of s-wave holographic superconductor phase transitions. The unordinary features might constitute 
the possible hints for future experiments testing the considered model of dark matter.
In the model in question, apart from the electromagnetic matter field we have taken into account 
the { \it dark matter} sector described by another $U(1)$-gauge field, bounded with 
the Maxwell field by the coupling constant $\alpha$. 

The models where dark matter is a part of a larger sector which interacts with visible matter were
successesfully implemented as the possible explanations of various astrophysical anomalous observations like
the excess of electrons in Galaxy having energies of a few GeV and TeV, gamma rays of 511 keV \cite{astro}. 
There were also efforts to find new physics explaining the anomalous muon magnetic moment, possible 
implication for parity violation, rare meson decays \cite{dav12}, as well as, to provide some implications
 of boson and {\it dark boson} mixing for high energy experiments \cite{dav13}. This problem is of a great 
importance especially in the light of the latest claim of nongravitational interactions
of dark matter in colliding galaxy clusters \cite{har15}, which can disfavor some extensions of the Standard Model.

In the paper we have discussed analytically various phase transitions toward 
the s-wave holographic superconductor in the probe limit. Our results  for $\alpha=0$
agree with the previous numerical and analytical studies \cite{cai11b,pan10} of holographic
superconductor transitions. The coupling between ordinary and { \it dark matter}
 changes the values of the parameters at the transition, however, only quantitatively for s-wave 
superconductors.

To make contact between transitions studied here and those known in the condensed matter 
systems, let us recall some basic facts from the latter field of research.
In the condensed matter systems there exist a number of metal/insulator transitions. 
They differ by the role played by the  interactions between carriers and the lack of periodicity of the 
underlying crystal lattice.   The metal/insulator transition may appear when the carriers strongly interact with each other. 
On the physical grounds it can be argued that in such strongly correlated system the electron movement is hindered
by the repulsive interactions and as a result, the insulator may form. The literature on the
Mott-Hubbard metal/insulator transition \cite{mott1990,hubbard1963} is vast and the transition 
itself is still not fully  understood \cite{imada1998}. The other interesting transition appears in the system which 
is not translationally invariant. The transition is driven by disorder. Strong disorder 
makes some states localized and thus unable to carry the electric current.  
This is called  Anderson metal/insulator  transition \cite{lee1985}. In real materials one usually finds
transitions in which both interactions and disorder do play an essential role \cite{belitz1994}.

Our study is restricted to the system which is periodic on the boundary and the insulator/metal transition we study 
should be related to the Mott-Hubbard one. The sequel of the transitions we are discussing, i.e., 
insulator/superconductor at zero temperature and metal/superconductor at higher temperatures are 
realized in high temperature cuprate and iron superconductors  with the increase of the carrier doping. 
The  $T=0$ insulator/superconductor transition is an analog of the Hawking - Page like  soliton-black 
hole transition.  In real superconductors increase of charge density beyond the upper limit induces
(at low temperatures) a reverse superconductor to metal transition, which seemingly has not been 
hitherto found in holographic analogy. 

Treating $\mu^2$ as a spectral parameter we analyze the behaviour of it 
in s-wave insulator/superconductor phase transition. We did not observe the influence of 
the {\it dark matter} sector on this quantity. The charge density  is proportional 
to the difference $(\mu - \mu_c)$, and also does not depend on the {\it dark matter} sector in the probe limit.
In the case of the scalar operator it was revealed that it is proportional to $(\mu - \mu_c)^{1/2}$
and represents the second order phase transition. The critical exponent of the considered system
has the mean-field value, while the proportionality factor is subject to {\it dark matter} coupling constant
dependence. The smaller value of $\alpha$ is considered, the greater factor one obtains, i.e., the harder condensation
happens (it will also be  the case in s-wave holographic metal/superconductor phase transition).

The  same conclusions were drawn studying s-wave type of the transition in question in Gauss-Bonnet theory. 
The Gauss-Bonnet coupling which  envisages the influence of higher curvature corrections, does influence 
the considered factor and the form of linear dependence survives.

In the case of holographic metal/superconductor phase transition, the critical temperature does
not depend on {\it dark matter} sector. The conclusion is in accord with our previous studies \cite{nak14},
where it was shown that the backreaction effects introduced the dependence of $T_c$ on {\it  dark matter} sector.
On the contrary, the condensation operator reveals a linear dependence on the $\alpha$-coupling constant. The 
bigger {\it dark matter} coupling constant one considers, the easier condensation forms.

Examining s-wave droplet insulator/superconductor phase transition, it was found that the chemical potential and 
magnetic field were bounded with the linear dependence. Not only does the magnetic field influence the condensation 
but also the coupling constant  of {\it dark matter} sector does this. The increase of the magnetic field
causes the increase of $\mu_c$, which eventuates in the harder condensation. For the fixed value of the magnetic field, 
it happens that
the smaller value of the {\it dark matter} sector coupling constant one chooses,
the harder condensation takes place. A very similar behaviour was envisaged in the holographic
droplet in p-wave insulator/superconductor phase transition case. 

To conclude, we remark that there are some points which are in contrast to the ordinary behaviour (without {\it
dark matter} sector) during the aforementioned phase transitions, which may constitute  indicators for the
future experiments for detecting dark matter and elucidating its nature.  
Testing s-wave holographic superconductors with {\it dark matter } sector is the only tip of the iceberg
and some more complicated models like p-wave or $p_x+ip_y$ should be taken into account. 
We hope to investigate these problems elsewhere. Our preliminary calculations show the stronger
modifications of the p-wave superconductor characteristics by the dark matter coupling $\alpha$.

\appendix
\section{Integration constants in Eq. (\ref{xi0})}
Equation (\ref{xi0}) provides formal solution of the differential equation (\ref{cc1}) and contains 
two integration constants. The value of $\xi(0)$ enters the prefactors $\Gamma$ and $\tilde{B}$ 
of the dependence of the condensation operator $<\cO_i>$ and the density $\rho$ on the chemical potential.

To find numerical values of the coefficients $\Gamma$ and $\tilde{ B}$ we note that the
integrals entering equation (\ref{xi0}) still contain terms singular at $z=1$ which have to be eliminated by
the proper choice of constants. To this end we take  $F(y)=1-ay^2$ and for numerical 
evaluation of the constants use the values of $a=a_{min}$ which
minimize the functional (\ref{funct-mu2}) for $\mu^2$. Evaluating the  (non-singular) 
integral over $y$ in Eq.(\ref{xi0}) leads to 
\ben
q^2~r_0^2\int_1^z dy ~y^{2\la_i -3}~F^2(y) 
= q^2~r_0^2 \left[y^{2\la_i}\left(-\frac{a}{\la_i}+\frac{1}{2(\la_j-1)y^2}+\frac{a^2y^2}{2(\la_i+1)}\right)\right]_1^z \nonumber \\
=q^2~r_0^2 \left(\left[z^{2\la_i}\left(-\frac{a}{\la_i}+\frac{1}{2(\la_j-1)z^2}+\frac{a^2z^2}{2(\la_i+1)}\right)\right]-
\left[\left(-\frac{a}{\la_i}+\frac{1}{2(\la_j-1)}+\frac{a^2}{2(\la_i+1)}\right)\right]\right).
\een
Inserting the above result denoted as $R(z)$ into (\ref{xi0}) and performing the integral over $z$ one gets
\be
\xi(0)=c_1-\int_0^1 dz \frac{z}{1-z^4}[c_2+R(z)] \nonumber \\
=c_1-\left[\frac{c_2}{4}(\log(1+z^2)-\log(1-z^2))+ q^2~r_0^2 W(z)\right]_0^1,
\ee
where we have denoted 
\ben
W(z)=\bigg[2\la_i(1+a^2(\la_i-1)+\la_i)z^{2\la_i+4}~_2F_1(1,1+\frac{\la_i}{2};2+\frac{\la_i}{2},z^4)-(2+\la_i)(-2z^{2\la_i}(\la_i+1) \nonumber \\
+4a(\la_i-1)z^{2\la_i+2}~_2F_1(1,\frac{1+\la_i}{2};\frac{3+\la_i}{2},z^4)+(a^2\la_i(\la_i-1)-\la_i(\la_i+1)-2a(\la_i^2-1))\log(1-z^2)
\nonumber \\
+(-2a+\la_i+a^2\la_i+\la_i^2+2a\la_i^2-a^2\la_i^2)\log(1+z^2))\bigg]/[8(\la_i-1)\la_i(\la_i+1)(\la_i+2)].
\een
Here $_2F_1(a,b;c,z)$ is the hypergeometric function \cite{abramowitz}. For the special values of parameters with $c=a+b$,
as in the above expression, it diverges for $z\rightarrow 1$ and takes the form \cite{abramowitz}
\be
\lim_{z\rightarrow 1}~ _2F_1(a,b;a+b,z)=-\frac{\Gamma(a+b)}{\Gamma(a)\Gamma(b)}\log(1-z).
\ee
It is easy to check that $W(0)=0$. On the other hand the boundary condition  $\xi(1)=0$ requires 
\be
c_2=-\frac{q^2~r_0^2}{\la_i-1},
\ee
and 
\be
c_1=-q^2~r_0^2\left[\frac{\log2}{4(\la_i-1)}+\frac{1}{4\la_i(\la_i-1)}+\left(\frac{a}{4\la_i(\la_i+2)}-\frac{a^2}{4(\la_i+1)(\la_i+2)}
-\frac{1}{8(\la_i-1)(\la_i+2)}\right)\log2\right].
\label{app-c1}
\ee

Introducing the obtained results into the equation(\ref{scaloper}), one gets the required formula
\be
<\cO_i>=\Gamma \sqrt{\mu-\mu_c},
\ee
where the prefactor $\Gamma=\sqrt{\frac{\tilde{\alpha}}{\mu_cq^2}}\sqrt{\frac{4(\lambda_i-1)}{(a-1)^2\log2}}$.
On the other hand, the charge density yields
\be
\rho=\frac{4(\lambda_i-1)\int_0^1{y^{2\lambda_i-3}F^2(y)dy}}{(a-1)^2\log2}(\mu-\mu_c).
\ee
Another way to find the required value $\xi(0)$ is by direct solution of the 
equation (\ref{cc1}).
For general value of the parameter $\lambda_i$ it is given in terms of the
hypergeometric $_2F_1(a,b;c,z)$ functions \cite{abramowitz} as
\ben
\xi(z)=C_2+\frac{q^2~r_0^2}{4(\lambda_i+2)(\lambda_i+1)\lambda_i(\lambda_i-1)}\left[\lambda_i(a^2(\lambda_i-1)
+\lambda_1+1)z^{2\lambda_i+4} {_2}F_1(1,\frac{\lambda_i}{2}+1;\frac{\lambda_i}{2}+2,z^4) \right. \nonumber \\
-\left. (\lambda_i+2)\left(2a(\lambda_i-1)z^{2\lambda_i+2} {_2}F_1(1,\frac{\lambda_i+1}{2};\frac{\lambda_i+3}{2},z^4)
+(\lambda_i+1)(-z^{2\lambda_i}+\lambda_i(\lambda_i-1)C_1\log\frac{1-z^2}{1+z^2} \right)\right].
\label{eq:sol-chi}
\een
Remembering that \cite{abramowitz} 
$\lim_{z\rightarrow 1}  {_2}F_1(1,\frac{\lambda_i}{2}+1;\frac{\lambda_i}{2}+2,z^4)=-(1+\frac{\lambda_i}{2})\log(1-z^2)$
and $\lim_{z\rightarrow 0}  {_2}F_1(1,\frac{\lambda_i}{2}+1;\frac{\lambda_i}{2}+2,z^4)=1$
one chooses the constants $C_1$ and $C_2$ in such a way that the solution is finite with $\chi(1)=0$ and
finds $\xi(0)=C_2$ with $C_2$ given by the formula (\ref{app-c1}) above.

\begin{acknowledgments}
{\L}N was supported by the Polish National Science Centre under FUGA $UMO-2014/12/S/ST2/00332$.\\
MR was partially supported by the grant of the National Science Center $DEC-2013/09/B/ST2/03455$
and KIW by the grant DEC-2014/13/B/ST3/04451.
\end{acknowledgments}



\end{document}